\documentclass[prb,aps,twocolumn,superscriptaddress,longbibliography]{revtex4-2}
\usepackage{graphicx,color}
\usepackage{amsthm}
\usepackage{amsfonts}
\usepackage{algorithmic}
\usepackage{enumerate}
\usepackage{latexsym}
\usepackage{amsmath}
\usepackage{amssymb}
\usepackage{multirow}
\usepackage{array}
\usepackage{diagbox}
\usepackage{subfigure}
\usepackage[colorlinks=true,citecolor=blue,linkcolor=blue]{hyperref}

\def\avg#1{\left\langle#1\right\rangle}

\emergencystretch=\maxdimen
\makeatletter
\DeclareFontFamily{OMX}{MnSymbolE}{}
\DeclareSymbolFont{MnLargeSymbols}{OMX}{MnSymbolE}{m}{n}
\SetSymbolFont{MnLargeSymbols}{bold}{OMX}{MnSymbolE}{b}{n}
\DeclareFontShape{OMX}{MnSymbolE}{m}{n}{
    <-6>  MnSymbolE5
   <6-7>  MnSymbolE6
   <7-8>  MnSymbolE7
   <8-9>  MnSymbolE8
   <9-10> MnSymbolE9
  <10-12> MnSymbolE10
  <12->   MnSymbolE12
}{}
\DeclareFontShape{OMX}{MnSymbolE}{b}{n}{
    <-6>  MnSymbolE-Bold5
   <6-7>  MnSymbolE-Bold6
   <7-8>  MnSymbolE-Bold7
   <8-9>  MnSymbolE-Bold8
   <9-10> MnSymbolE-Bold9
  <10-12> MnSymbolE-Bold10
  <12->   MnSymbolE-Bold12
}{}

\let\llangle\@undefined
\let\rrangle\@undefined
\DeclareMathDelimiter{\llangle}{\mathopen}%
                     {MnLargeSymbols}{'164}{MnLargeSymbols}{'164}
\DeclareMathDelimiter{\rrangle}{\mathclose}%
                     {MnLargeSymbols}{'171}{MnLargeSymbols}{'171}
\makeatother

\NewDocumentCommand{\dgal}{sO{}m}{%
  \IfBooleanTF{#1}
    {\dgalext{#3}}
    {\dgalx[#2]{#3}}%
}
\NewDocumentCommand{\dgalext}{m}{%
  \sbox0{%
    \mathsurround=0pt 
    $\left\{\vphantom{#1}\right.\kern-\nulldelimiterspace$%
  }%
  \sbox2{\{}%
  \ifdim\ht0=\ht2
    \{\kern-.625\wd2 \{#1\}\kern-.625\wd2 \}%
  \else
    \left\{\kern-.7\wd0\left\{#1\right\}\kern-.7\wd0\right\}%
  \fi
}

\begin{document}
\title{Competition between $d$-wave and $d$+$is$-wave superconductivity in the Hubbard model on a checkerboard lattice}
\author{Yue Pan}
\affiliation{School of Physics and Astronomy, Beijing Normal University, Beijing 100875, China\\}
\author{Runyu Ma}
\affiliation{School of Physics and Astronomy, Beijing Normal University, Beijing 100875, China\\}
\author{Chao Chen}
\affiliation{Department of Basic Courses, Naval University of Engineering, Wuhan 430033, China\\}
\affiliation{School of Physics and Astronomy, Beijing Normal University, Beijing 100875, China\\}
\affiliation{Beijing Computational Science Research Center, Beijing 100193, China}
\author{Zixuan Jia}
\affiliation{School of Physics and Astronomy, Beijing Normal University, Beijing 100875, China\\}
\author{Tianxing Ma}
\email{txma@bnu.edu.cn}
\affiliation{School of Physics and Astronomy, Beijing Normal University, Beijing 100875, China\\}
\affiliation{Key Laboratory of Multiscale Spin Physics (Ministry of Education), Beijing Normal University, Beijing 100875, China}

\begin{abstract}
By employing determinant quantum Monte Carlo simulations,
we investigate a checkerboard lattice with next-nearest-neighbor hopping $t'$ as the
frustration-control parameter, which exhibits an energetically partial flat-band in the system.
Our numerical simulation identifies the dominant pairing symmetry of the checkerboard lattice Hubbard model,
and we reveal the competition between the $d$-wave and $d+is$ wave
in the parameter space of electron filling $\avg{n}$ and frustration control parameter $t^{\prime}/t$.
To ensure the reliability and accuracy of our results, we evaluate the sign problem.
We also find that the spin susceptibility, the effective pairing interactions of different pairing symmetries and the
superconducting instability
are enhanced as the on-site Coulomb interaction increases,
demonstrating that superconductivity is driven by strong electron--electron correlation.
Our work provides a further understanding of pairing symmetry in the Hubbard model 
and improves prospects for exploring rich correlated behaviors in frustrated systems.
\end{abstract}
\maketitle

\section{Introduction}
In strongly correlated system, exploring the physics of superconductivity has been a significant research topic for decades.
From the theoretical viewpoint, it is believed that Hubbard model on a square lattice is closely related to the physics of high-$T_{c}$ superconductivity in doped cuprates\cite{Bednorz1986,CHU2015290,Keimer2015}.
After extensive studies by different methods,
it is found that some additional terms added in the original Hubbard model, or changing the geometry of the lattice, may play an important role in the shaping of superconducting phase\cite{hubbard1967electron,PhysRevLett.95.237001,PhysRevB.81.224505}.
A series of studies have found that partial flat bands contribute significantly to the formation of superconductivity in different lattice geometries.
For example, twisted bilayer graphene\cite{andrei2020graphene,PhysRevX.10.031034} and transition
metal dichalcogenides \cite{wang2020correlated,zhang2020flat} have provided a strong motivation for studying unconventional superconductivity \cite{PhysRevB.99.121407,balents2020superconductivity,huang2019antiferromagnetically}, in which the isolated electronic flat band appearing at the magical angle plays an essential role.
Additionally, the interplay of frustration and superconductivity may be another key
concept in the investigation of superconductors.
Recently, many studies have emphasized the effect of next-nearest-neighbor (NNN) hopping $t^{\prime}$ in $t$-$t^{\prime}$-$J$ model,
where $t$ represents the nearest hopping (NN)
\cite{doi:10.1073/pnas.2109978118, xu2023coexistence}.
The additional $t^{\prime}$
term accommodates the particle-hole asymmetry,
introducing frustration helps stabilize superconductivity by suppressing stripe order
\cite{zhang2023frustrationinduced,PhysRevB.75.184523,science.aal5304}.
The checkerboard lattice, which describes the physics of pyrochlore oxides hosting superconductivity and other 
strong electronic correlated phenomenon \cite{balents2010spin,PhysRevLett.89.226402,PhysRevLett.120.196401}, is a multiband system involving a tunable flat band and frustration, sparking our research interest.

Based on above considerations, we conducted our investigations on a checkerboard lattice Hubbard model.
By tuning the hopping parameters, we can easily adjust the range of the flat band in the momentum space in checkerboard lattice \cite{PhysRevB.107.245126,PhysRevB.78.165113}. 
This striking property motivated us to study pairing symmetries in the checkerboard lattice Hubbard model. 
Materials based on checkerboard lattices have been predicted to host exotic quantum physics.  
The actual band structure of these materials is considerably 
more complex and includes a series of 
pyrochlore oxides,
but their important features are
believed to originate from their two-dimensional(2D) checkerboard lattice substructure, two atoms in each unit cell.
Examples include the quantum spin liquid candidate $\mathrm{Cd}_2\mathrm{Zr}_2\mathrm{O}_7$ \cite{PhysRevLett.122.187201,gao2019experimental,bhardwaj2022sleuthing}
and superconductor $\mathrm{Cd}_2\mathrm{Re}_2\mathrm{O}_7$ \cite{PhysRevB.101.045117,PhysRevResearch.2.033108}. 
The latter has drawn attention since the discovery of superconductivity in 2001 with critical temperature $T_c\approx 1.0 K$,
and it has added a new family to the growing class of materials with possible unconventional superconductivity \cite{PhysRevLett.87.187001,PhysRevB.64.180503,sakai2001superconductivity,santos2010two,wu2019possible}.
Recently, a related investigation found some exotic states of pairing symmetry on a checkerboard lattice by renormalized mean-field theory \cite{PhysRevB.75.184523} and finite cluster 
exact diagonalization study \cite{PhysRevLett.93.197204}. 
Concurrently, advances have been made possible both by
improvements in measurement techniques and by the discovery of entirely new classes of checkerboard lattice material.
It was reported that theoretical prediction and experimental realization in monolayer $\mathrm{Cu}_2\mathrm{N}$, 
showed the flat band and
saddle point near the Fermi
level, which might give rise to intriguing properties and was crucial for further device applications \cite{doi:10.1021/acs.nanolett.3c01111}.

Here, we use the determinant quantum Monte Carlo (DQMC) method to study the Hubbard model on a checkerboard lattice 
with the presence of NNN $t^{\prime}$, in which doping and on-site Coulomb interactions are also considered.
For a wide range of key parameters, 
we discuss the dominant pairing symmetries in this system.
It is shown that the $d$-wave and $d$+$is$ wave are dominant over other pairing symmetries 
in the parameter space of electron filling $\avg{n}$ and frustration control parameter $t^{\prime}/t$.
It is also found that the spin susceptibility, the effective pairing interaction of different pairing symmetries and superconducting instability 
are enhanced as the interaction strength increases,
demonstrating that the superconductivity is driven by strong electron-electron correlation.
The DQMC method is a numerically exact and unbiased approach used to simulate quantum many-body
models. The fermion sign
problem usually exists in general, and the accessible temperatures and interaction strength are restricted.
However, this can probably be circumvented by calculating much longer runs,
and we evaluate the sign problem to clarify which parameter regions are accessible and reliable.
Our systematic DQMC results provide new insights into
the problem of pairing symmetry in a flat-band system.

\section{Model and methods}
Our starting point is the Hubbard model on
a checkerboard lattice.
With the introduction of hopping inhomogeneity, this
model Hamiltonian can be written as:
\begin{equation}
\begin{aligned}
{H}=&-t\sum_{\langle i,j \rangle \sigma}(a_{i\sigma}^\dagger b_{j\sigma}+b_{j\sigma}^{\dagger}a_{i\sigma})\\
    &-t^{\prime}\sum_{ \ll i,j \gg \sigma}(a^{\dagger}_{i\sigma}a_{j\sigma}+{a}^{\dagger}_{j\sigma} a_{i\sigma}) \\
    &-t^{\prime}\sum_{\dgal*{i,j} \sigma}(b^{\dagger}_{i\sigma}b_{j\sigma}+{b}^{\dagger}_{j\sigma} b_{i\sigma})\\
    &+U \sum_{i} (n_{iA\uparrow} n_{iA\downarrow}+ n_{iB\uparrow} n_{iB\downarrow})\\
    &+\mu \sum_{i\sigma} (n_{iA\sigma}+n_{iB\sigma})
\end{aligned}
\end{equation}
in which $a^{\dagger}_{i\sigma}$ ($a_{i\sigma}$) is the creation (destruction) operator for an
electron of spin $\sigma$ on lattice site $i$ at A sublattice, and $b^{\dagger}_{i\sigma}$ ($b_{i\sigma}$)
is the counterpart on B sublattice.
$U$ is the on-site repulsive
interaction, and $\mu$ is the chemical potential used to tune the electron filling. 
$\langle i,j \rangle$ denotes NN sites, and they are connected by hopping $t$,
while $\ll i, j \gg $ denotes NNN sites of sublattice A where $\mathbf{R}_{i} - \mathbf{R}_{j} = \pm(1,-1)$
and $\dgal*{i, j} $ denotes NNN sites of sublattice B where $\mathbf{R}_{i} - \mathbf{R}_{j} = \pm (1,1)$.
In this work, we set $t=1$ as the energy scale.

We use the DQMC algorithm \cite{PhysRevD.24.2278,PhysRevB.63.125116} to study the pairing symmetries of this system. 
In the DQMC method, the expectation values of observables $\langle M \rangle =\text{Tr} M\exp(-\beta H)/\text{Tr}\exp(-\beta H)$ 
are evaluated by discretizing the inverse temperature $\beta$ and integrating out the fermion
degrees of freedom. To address the interaction terms in the Hamiltonian, one needs to perform a Hubbard--Stratonovich (HS) transformation to write the interaction term into the quadratic term. 
Then, the partition
function can be converted into the product of two fermion
determinants, where one is spin up and the other is spin down.
We can measure the severity of the sign problem
by computing the ratio of the integral of the product of up and down spin determinants, to the integral of the
absolute value of the product.
The average 
sign $\langle sign \rangle$ is written as\cite{PhysRevB.41.9301,PhysRevB.92.045110}
\begin{equation}
\begin{aligned}
\langle\mathrm{sign}\rangle=\frac{\sum_{\mathcal{X}}\mathrm{det}M_{\uparrow}(\mathcal{X})\det M_{\downarrow}(\mathcal{X})}{\sum_{\mathcal{X}}|\mathrm{det}M_{\uparrow}(\mathcal{X})\det M_{\downarrow}(\mathcal{X})|},
\end{aligned}
\end{equation}
Here, $M_{\sigma}(\mathcal{X})$ represents each spin specie matrix. 
The $\langle sign \rangle$ equals 
to 1 indicates the absence of sign problem.
In the simulations, we use 8000 sweeps to equilibrate the system and an
additional 10000$\sim$200000 sweeps to generate measurements. These measurements are split into 10 bins and provide the basis of the
coarse-grain averages. The errors are calculated based on the standard deviation from the average. For more technical details, 
please see Refs. \cite{DQMC,PhysRevB.28.4059}.

\begin{figure}[tbp]
\includegraphics[scale=0.34]{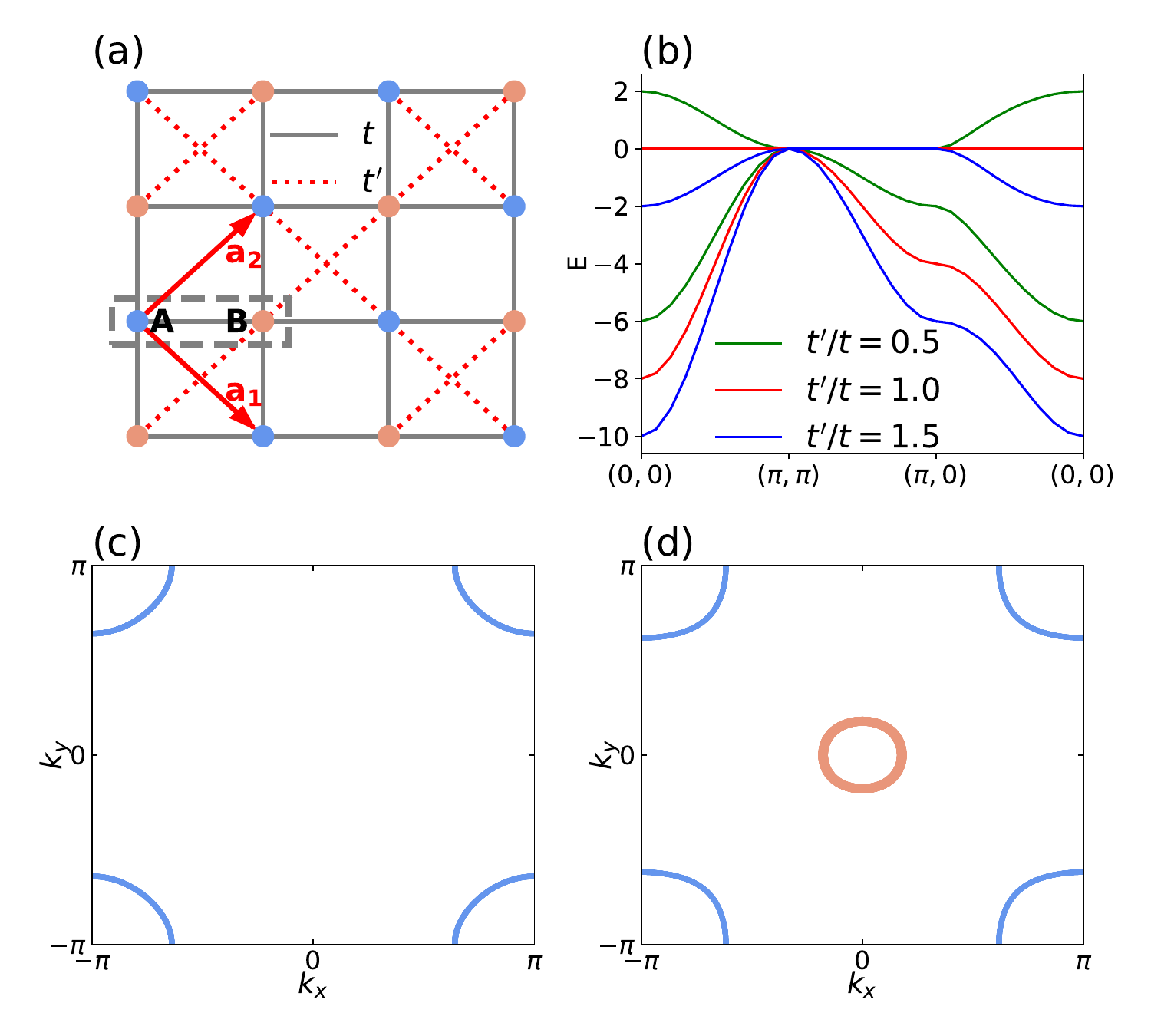}
\caption{(Color online) (a) Sketch of a checkerboard lattice, (b) the energy band along the high-symmetry
line in the unfolded Brillouin zone, (c) the Fermi surface at $t^{\prime}/t=1.0$, $\avg{n}=0.9$, and (d) $t^{\prime}/t=1.5$, $\avg{n}=0.9$.
}
\label{Fig1}
\end{figure}

\section{Results and discussion}
The unit vector in the checkerboard lattice is shown in Fig. \ref{Fig1}(a), with the two
nonequivalent sites A and B in the unit cell. 
Fig. \ref{Fig1}(b) displays the band structure along with the
high-symmetry path of the first Brillouin zone for different $t^{\prime}/t$, 
which reflects the character of these two sublattices, showing that there is an energetically 
flat band.
We plot the Fermi surface at $t^{\prime}/t=1.0$ and $t^{\prime}/t=1.5$ in Fig. \ref{Fig1}(c) and (d).  
A subtle but significant change in the Fermi
surface may take place at the transition, where a second closed Fermi surface emerges at $t^{\prime}/t = 1.5$.

To study magnetic properties, we define the spin susceptibility in the $z$ direction at zero frequency,
\begin{equation}
\chi(q) = \frac{1}{N} \sum_{i, j} \sum_{l, m=A/B} \int^{\beta}_{0} d \tau e^{-i q (\mathbf{R}_{i} - \mathbf{R}_{j})}
\left\langle S^{z}_{il}(\tau) S^{z}_{jm}(0) \right\rangle
\label{eq2}
\end{equation}
where $N=2 \times L^2$ is the number of sites.
$S^{z}_{il}(\tau)=e^{H \tau }S^{z}_{il}(0)e^{-H\tau}$ with
$S^{z}_{iA}=a^{\dagger}_{i\uparrow} a_{i\uparrow}-a^{\dagger}_{i\downarrow}a_{i\downarrow}$, and
$S^{z}_{iB}=b^{\dagger}_{i\uparrow} b_{i\uparrow}-b^{\dagger}_{i\downarrow}b_{i\downarrow}$.
To gain a deep understanding of the effect of the next-nearest-neighbor hopping $t^{\prime}$ as 
frustration-control parameter on the magnetic order, we show the
magnetic susceptibility as a function of temperature $T$, interaction $U$ and electron filling $\avg{n}$ for several typical values $t^{\prime}/t = 0.25 \sim 1.50$.
The total number of lattice sites we focus on in this paper is $N = 2\times{L^2}$ with $L = 8$.

\begin{figure}[tbp]
\includegraphics[scale=0.4]{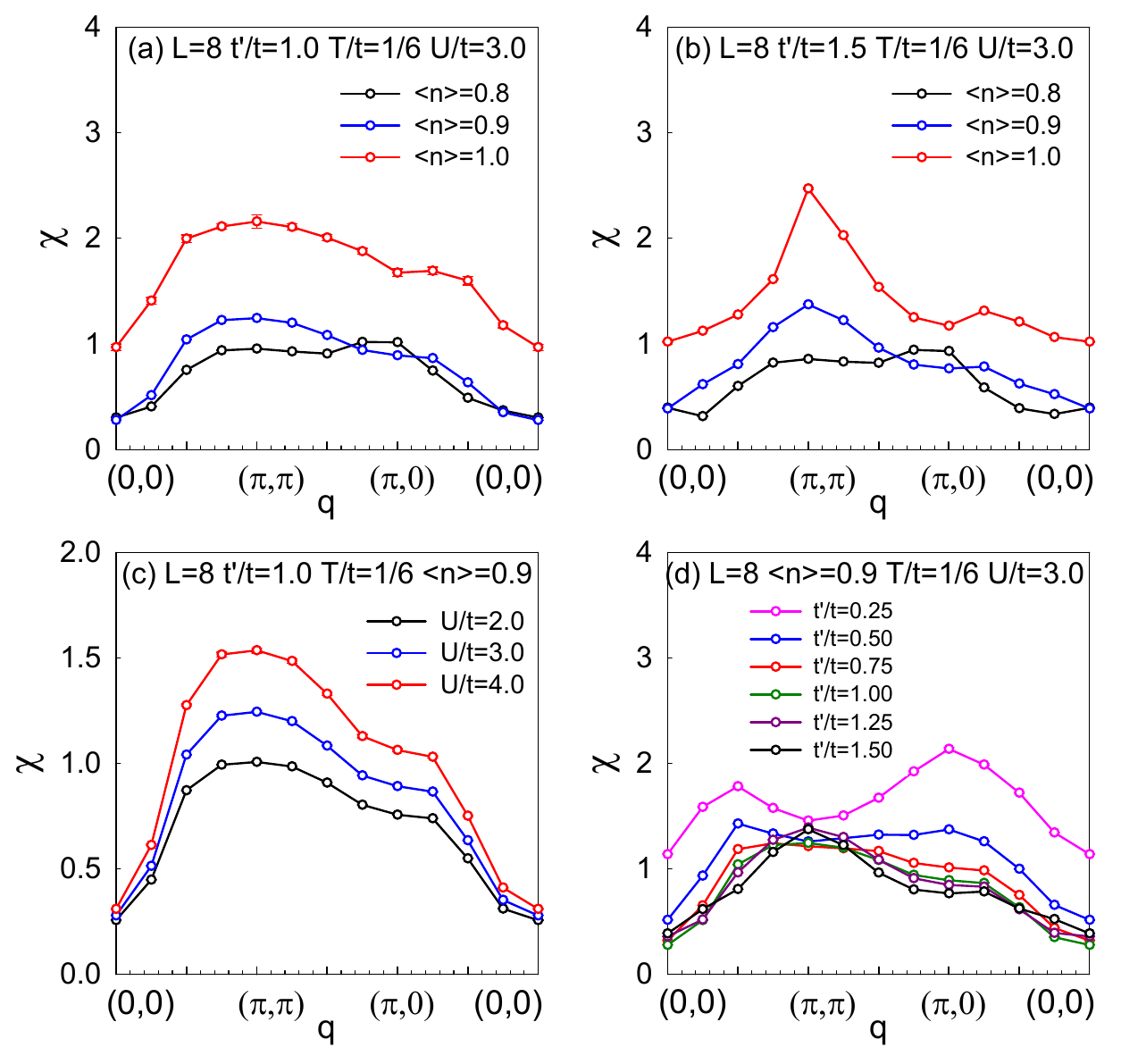}
\caption{(Color online) The magnetic susceptibility versus momentum $q$ at different values of $\avg{n}$ with $L = 8$, $T/t = 1/6$, $U/t = 3.0$ (a) $t^{\prime}/t=1.0$, and (b) $t^{\prime}/t=1.5$. (c) The magnetic susceptibility versus momentum $q$ at
different values of $U$ with $L = 8$, $t^{\prime}/t=1.0$, $T/t= 1/6$ and $\avg{n}=0.9$. 
(d) The magnetic susceptibility versus momentum $q$ at
different values of $t^{\prime}/t$ with $L = 8$, $\avg{n}=0.9$, $T/t = 1/6$ and $U/t = 3.0$.
}
\label{Fig2}
\end{figure}

In Fig. \ref{Fig2}(a)-(d), we display the magnetic susceptibility along the
high-symmetry path of the first Brillouin zone.
We calculate the magnetic susceptibility via Eq. \eqref{eq2}, which
depends on the momentum transfer $q$, for increasing values of $\avg{n}$ and
the same values of $T/t = 1/6$ and $U/t=3.0$ in Fig. \ref{Fig2}(a) $t^{\prime}/t=1.0$ and Fig. \ref{Fig2}(b) $t^{\prime}/t=1.5$.
The magnetic susceptibility increases as $\avg{n}$ increases.
We also find that the maximum value of magnetic susceptibility $\chi$ 
at $q = (\pi,\pi)$ is as expected for the repulsive Hubbard model
on a checkerboard lattice, which is consistent with the scenario that
the antiferromagnetic fluctuations mediate the pairing on the square lattice. 
In Fig. \ref{Fig2}(c), we show the magnetic susceptibility at electron filling $\avg{n}=0.9$ and
fixed $t^{\prime}/t = 1.0$ for increasing values of $U$.
We know that the value of the magnetic susceptibility increases as the on-site Coulomb interaction $U$ increases.
Similarly, we show the magnetic susceptibility for increasing values $t^{\prime}/t$ from 0.25 to 1.5 
in Fig. \ref{Fig2}(d). As $t^{\prime}/t$ increases, the value of the magnetic susceptibility decreases.

To study the superconducting property of the checkerboard lattice, we calculate the pairing susceptibility\cite{PhysRevLett.110.107002},
\begin{equation}
  P_{\alpha }=\frac{1}{N}\sum_{i,j} \int_{0}^{\beta }d\tau \langle \Delta
  _{\alpha }^{\dagger }(i,\tau)\Delta_{\alpha }^{\phantom{\dagger}}(j,0)\rangle
  \label{shi3}
\end{equation}
where $\alpha $ denotes the pairing symmetries. 
In this work, we mainly investigate the singlet pairings on NN bonds, and
the corresponding order parameter $\Delta_{\alpha }^{\dagger}(i,\tau)=e^{H \tau }\Delta_{\alpha }^{\dagger}(i,0)e^{-H\tau}$, 
in which $\Delta_{\alpha }^{\dagger}(i)=\Delta_{\alpha }^{\dagger}(i,0)$\ is written as
\begin{equation}
  \begin{aligned}
    \Delta _{\alpha }^{\dagger }(j)=
    \sum_{I}g_{I}f_{\alpha }^{\dagger }(j,\delta_{I})
    \begin{cases} 
      (a_{j\uparrow }a_{{j+{\delta_{I}}}\downarrow }-a_{{j}\downarrow }a_{{j+\delta_{I}}\uparrow })^{\dagger } \mbox{if } j \in A \\
      (b_{j\uparrow }b_{{j+{\delta_{I}}}\downarrow }-b_{{j}\downarrow }b_{{j+\delta_{I}}\uparrow })^{\dagger } ~\mbox{if } j \in B
    \end{cases}
  \end{aligned}
\label{eq4}
\end{equation}
where $g_{I}f_{\alpha}(j,\delta_I)$ stands for the form factor of the pairing function. The coefficient $g_{I} = \sqrt{3}/2$ for $I \leq 4$ and  $g_{I} = 1/2$ for $I=5,6$. 

By averaging the single particle Green functions  before taking their product, the interaction vertex is removed.
This consists of two dressed Green functions without mutual correlations. We call this quantity the uncorrelated susceptibility $\widetilde{P}_{\alpha}$.
This is used to help us determine whether the interaction between the dressed quasi-particle is attractive or repulsive.
$\widetilde{P}_{\alpha}$ is calculated by replacing the four fermion terms $\langle c^{\dagger}_{\uparrow} c^{\dagger}_{\downarrow} c_{\downarrow} c_{\uparrow} \rangle$ with
$\langle c^{\dagger}_{\uparrow} c_{\uparrow} \rangle \langle c^{\dagger}_{\downarrow} c_{\downarrow} \rangle$.
We define the effective pairing interaction ${\rm{P}_\alpha}= P_{\alpha} - \widetilde{P}_\alpha$, and
the interaction vertex $\Gamma_\alpha$ is extracted as\cite{PhysRevB.39.839}
\begin{equation}
\Gamma_\alpha=\dfrac{1}{P_\alpha}-\dfrac{1}{\widetilde{P}_\alpha}.
\end{equation}
Eq. (4) can be rewritten as
\begin{equation}
P_\alpha=\dfrac{\widetilde P_\alpha}{1+\Gamma_\alpha\widetilde P_\alpha},
\end{equation}
and this suggests that $\Gamma_\alpha\widetilde P_\alpha \rightarrow -1$ signals superconducting instability. 

\begin{figure}[tbp]
\includegraphics[scale=0.8]{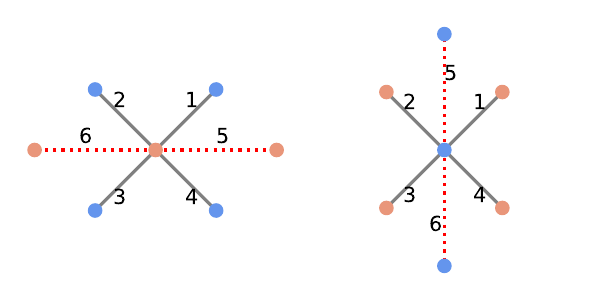}
\caption{(Color online) Phases of the pairing symmetries on the checkerboard lattice. 
Here, the different-colored dots denote the sites of the different sublattices A and B.}
\label{Fig3}
\end{figure}   

In Eq. \eqref{eq4}, vectors $\delta_I$ denote NN intersublattice connections and two NNN intrasublattice connections, 
as sketched in Fig. \ref{Fig3}.
Considering the symmetry of checkerboard lattice, and following the pairing forms shown in Ref.\cite{PhysRevB.75.184523},
the four form factors of $s$-wave, $d$-wave, $d+id$ wave, and $d$+$is$ wave pairings are given by 
\begin{equation}
  \begin{aligned}
  f_{s}(j,\delta_I) &= \begin{cases}1 & I\le4\\ -1 & I=5,6\end{cases}  \\
  f_{d}(j,\delta_I) &= \begin{cases} (-1)^I & I\leq4\\ 0 & I=5,6 \end{cases} \\
  f_{d+id}(j,\delta_{I}) & =\begin{cases} (-1)^{I} & I\leq4 \\  -i & I=5,6 ~\mbox{if } j \in A \\ i & I=5,6 ~\mbox{if } j \in B\\ \end{cases}  \\
  f_{d+is}(j,\delta_I) & =\begin{cases}(-1)^I & I\le4 \\ i  & I=5,6 \end{cases}\emph{} \\
  \end{aligned}
  \end{equation}
Here the defined $s$-wave is the same as $s$-$s$ wave in Ref. \cite{PhysRevB.75.184523}.

\begin{figure}[tbp]
\includegraphics[scale=0.35]{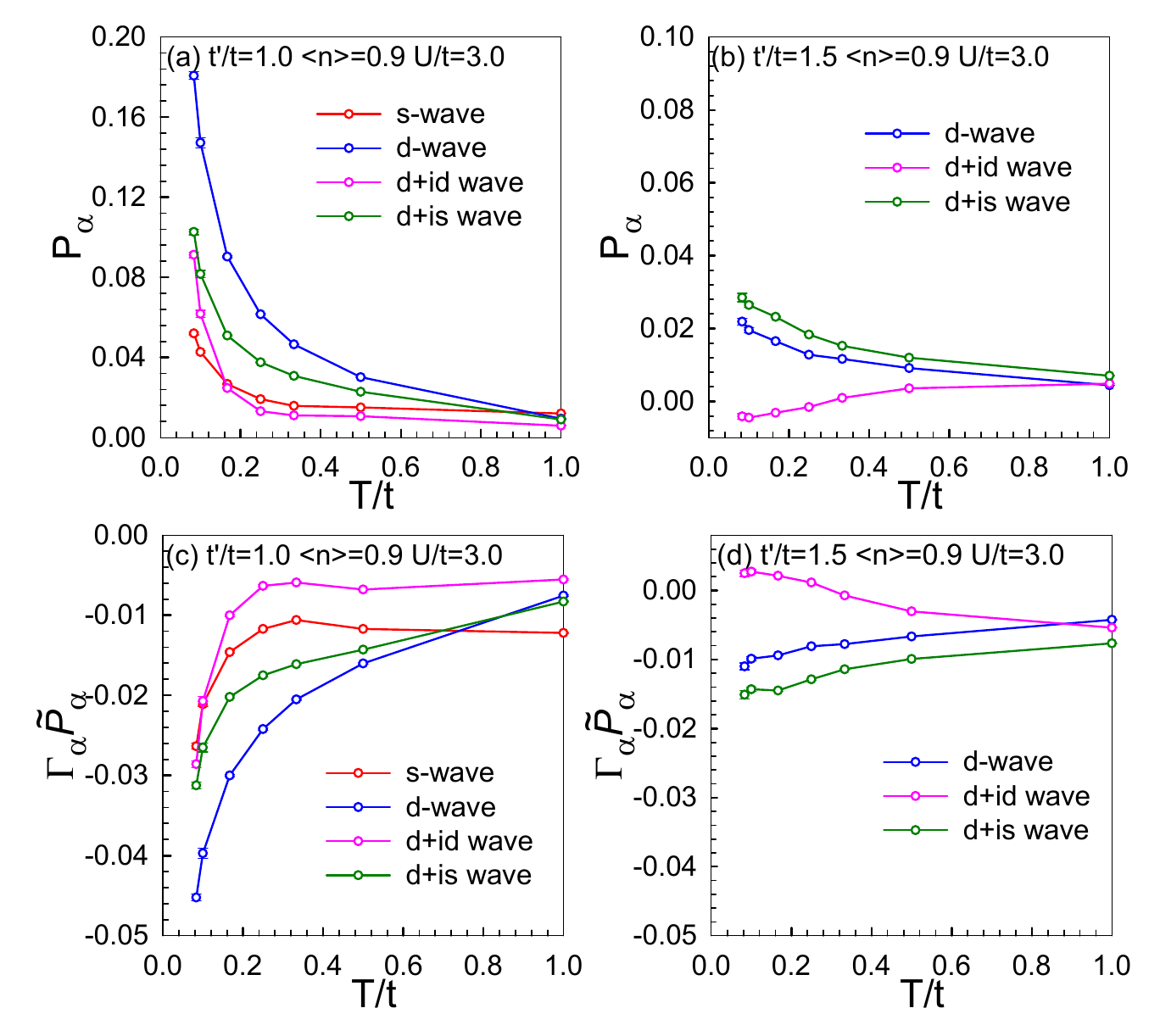}
\caption{(Color online) The effective pairing interaction $\rm{P}_\alpha$ of different pairing symmetries as a function of temperature
at (a) $t^{\prime}/t=1.0$ and (b) $t^{\prime}/t=1.5$ with electron
filling $\avg{n}=0.9$ and interaction strength $U/t = 3.0$ on a $2\times8^2$ lattice.
$\Gamma_\alpha\widetilde P_\alpha$ of different pairing symmetries as a function of temperature 
at (c) $t^{\prime}/t=1.0$ and (d) $t^{\prime}/t=1.5$ with electron
filling $\avg{n}=0.9$ and interaction strength $U/t = 3.0$ on a $2\times8^2$ lattice.
}
\label{Fig4}
\end{figure}

First, we examine the temperature dependence of $\rm{P}_\alpha$ for
different pairing symmetries in the
case of $U/t=3.0$ and filling $\avg{n}=0.9$ for $t^{\prime}/t=1.0$ and $t^{\prime}/t=1.5$.
In Fig. \ref{Fig4}(a), it is clearly observed that the effective pairing interaction of different pairing symmetries increases with decreasing temperature,
especially the $d$-wave pairing symmetry, where the effect
of temperature on $\rm{P}_\alpha$ is more pronounced than other pairing symmetries.
This indicates that the $d$-wave pairing symmetry is the dominant pairing symmetry at $t^{\prime}/t=1.0$.
However, in Fig. \ref{Fig4}(b), the $d$+$is$ wave pairing symmetry is dominant over other pairing symmetries at $t^{\prime}/t=1.5$.
The emergence of the $d$+$is$ wave is accompanied by the deformation of the Fermi surface.
When $t^{\prime}$ exceeds $t$, $t^{\prime}/t>1$, the energy of the upper band decreases around $q = (0, 0)$ point, and the upper band will also intersect the Fermi level.
Then, two closed Fermi surfaces appear, as shown in Fig. \ref{Fig1}(c) and (d).
We examine interaction vertex $\Gamma_\alpha\widetilde P_\alpha$ with different pairing symmetries, 
signaling a superconducting instability 
by $\Gamma_\alpha\widetilde P_\alpha \rightarrow -1$. 
The amplitude of dominant pairing symmetry increases as the temperature
decreases in Fig. \ref{Fig4}(c) and (d). 

\begin{figure}[tbp]
\includegraphics[scale=0.35]{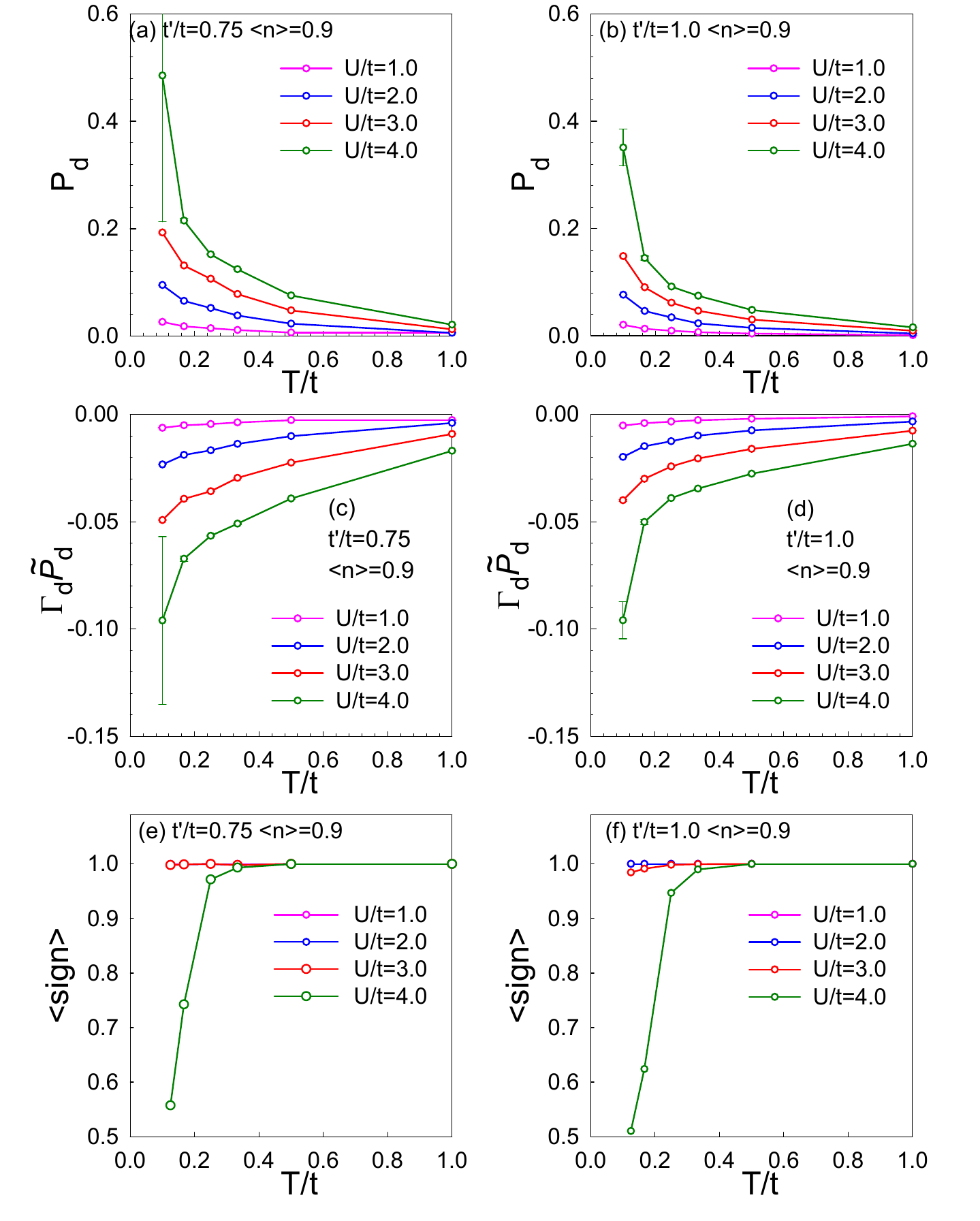}
\caption{(Color online) The $d$-wave effective pairing interaction $\rm{P}_d$ as a function of temperature for different $U$ at
(a) $t^{\prime}/t=0.75$ and (b) $t^{\prime}/t=1.0$ with $\avg{n}=0.9$ on a $2\times8^2$ lattice.
$\Gamma_d\widetilde P_d$ as a function of temperature for different $U$ at (c) $t^{\prime}/t=0.75$ 
and (d) $t^{\prime}/t=1.0$ with $\avg{n}=0.9$ on a $2\times8^2$ lattice.
The average sign as a function of temperature $T$ for
different $U$, $\avg{n}=0.9$ at (e) $t^{\prime}/t=0.75$ and (f) $t^{\prime}/t=1.0$. 
}
\label{Fig5}
\end{figure}
  
On the basis of above results, we plot the temperature dependence of $d$-wave effective pairing interaction $\rm{P}_d$ under a few representative values $U/t=1.0\sim4.0$ in Fig. \ref{Fig5}(a) $t^{\prime}/t=0.75$ and Fig. \ref{Fig5}(b)$t^{\prime}/t=1.0$ at filling $\avg{n}=0.9$. 
These further results show that the $d$-wave pairing symmetry is enhanced greatly as the value of
$U$ increases, which indicates that the strong
electron-electron correlation plays a key role in driving superconductivity.
In particular, our results illustrate that the qualitative evolution
of $d$-wave pairing is robust at different interaction strengths $U$,
as has been verified in many previous studies for different systems\cite{PhysRevB.90.075121,PhysRevB.106.195112,PhysRevB.104.035104,PhysRevLett.110.107002}.
We also examine the effect of interaction on vertex $\Gamma_d\widetilde P_d$ in Fig. \ref{Fig5}(c) $t^{\prime}/t=0.75$ 
and Fig. \ref{Fig5}(d) $t^{\prime}/t=1.0$. 
$\Gamma_d\widetilde P_d$ decreases as the interaction strength increases from $U/t= 1.0$ to $U/t = 4.0$, which are
consistent with Fig. \ref{Fig5}(a) and (b), indicating that the $d$-wave effective pairing interaction is significantly
dependent $U$.  We also compute the relation between $\langle sign \rangle$ and $T$ in Fig. \ref{Fig5}(e) for $t^{\prime}/t=0.75$ and 
Fig. \ref{Fig5}(f) 
for $t^{\prime}/t=1.0$.
From the numerical results at $\avg{n}=0.9$ and $U = 1.0t\sim 4.0t$, the average sign decays exponentially with
decreasing temperature. 
The average sign is in the vicinity of 1 when the temperature is high, providing the reliable 
numerical data; with the decrease of temperature, the value of $\langle sign \rangle$
begins to reduce, indicating that the sign problem becomes worse.
When $\langle sign \rangle$ is larger than 0.5, the sign problem is mild 
and does not prohibit obtaining accurate results.  
For the results where the sign problem is serious, longer runs can help compensate for the fluctuations by the DQMC method. 
Thus, we additionally compute with Monte Carlo parameters of much longer runs as $\langle sign \rangle$ is smaller than 0.5.

\begin{figure}[tbp]
\includegraphics[scale=0.35]{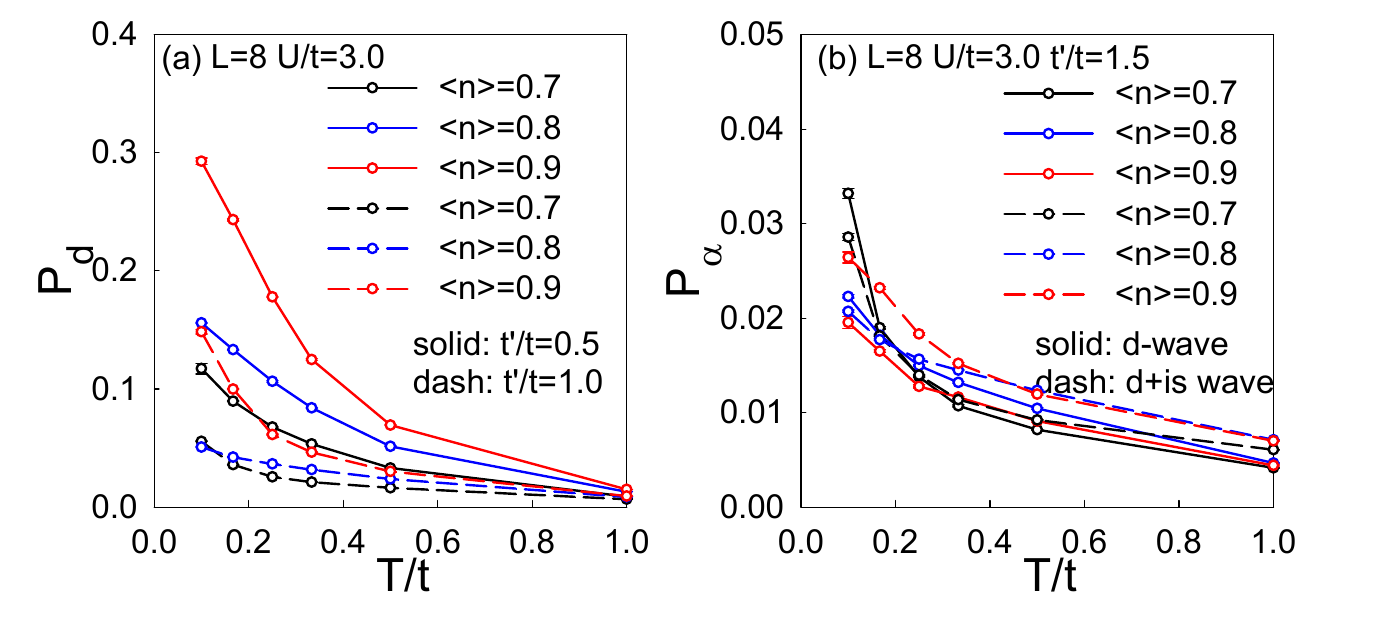}
\caption{(Color online) 
(a) The $d$-wave effective pairing interaction $\rm{P}_d$ as a function of temperature for different $\avg{n}$ and $U/t=3.0$ at $t^{\prime}/t=0.50$ (solid line) 
and $t^{\prime}/t=1.0$ (dashed line) on a $2\times8^2$ lattice.
(b) The effective pairing interaction $\rm{P}_\alpha$ as a function of temperature with different $\avg{n}$, $U/t=3.0$ and $t^{\prime}/t=1.5$ for $d$-wave (solid line) 
and $d$+$is$ wave (dashed line) on a $2\times8^2$ lattice. 
}
\label{Fig6}
\end{figure}

\begin{figure}[tbp]
\includegraphics[scale=0.45]{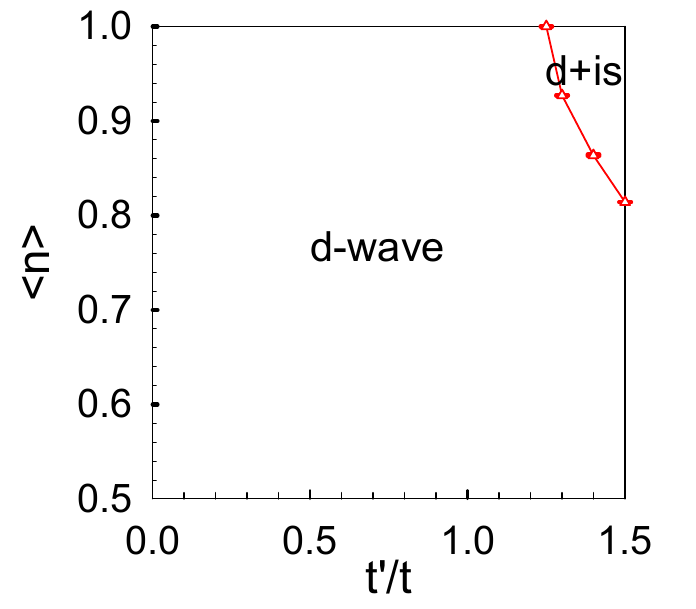}
\caption{(Color online) Dominant pairing symmetry on a checkerboard 
lattice in the parameter space of electron filling $\avg{n}$ and frustration control parameter $t^{\prime}/t$.
}
\label{Fig7}
\end{figure}

To describe the role of doping, we investigate the $d$-wave effective pairing interaction $\rm{P}_d$ 
with different doping $\langle{n}\rangle = 0.7$, 0.8, and 0.9 at $U/t=3.0$ and fixed 
$t^{\prime}/t$ strength, as shown in Fig. \ref{Fig6}(a) and (b). On the one hand, increasing the electron filling will enhance 
the effective pairing interaction. 
On the other hand, comparing $t^{\prime}/t=0.5$ (solid line) and $t^{\prime}/t=1.0$ (dashed line) at the same filling, 
one can see that $\rm{P}_d$ decreases as the next-nearest-neighbor
$t^{\prime}/t$ increases.
To compare the $d$-wave (solid line) and $d$+$is$ wave (dashed line) at $t^{\prime}/t=1.5$, 
dominant pairing symmetry at relatively low temperatures from $\langle{n}\rangle = 0.7$ to $\avg{n}=0.9$ 
is given in Fig. \ref{Fig6}(b). 
To have a global picture, we plot the dominant pairing symmetry on a checkerboard
lattice in the region of frustration control parameter $t^{\prime}/t$ in Fig. \ref{Fig7}. 
The phase diagram reveals that $d$-wave and $d$+$is$-wave are two possible dominant pairings in the parameter space
of electron filling $\avg{n}$ and frustration control parameter $t^{\prime}/t$. 
Remarkably, the $d$-wave
pairing has been previously found to be stable against
weak frustrations by introducing NNN hopping\cite{science.aal5304,PhysRevB.75.184523}.

\begin{figure}[tbp]
\includegraphics[scale=0.35]{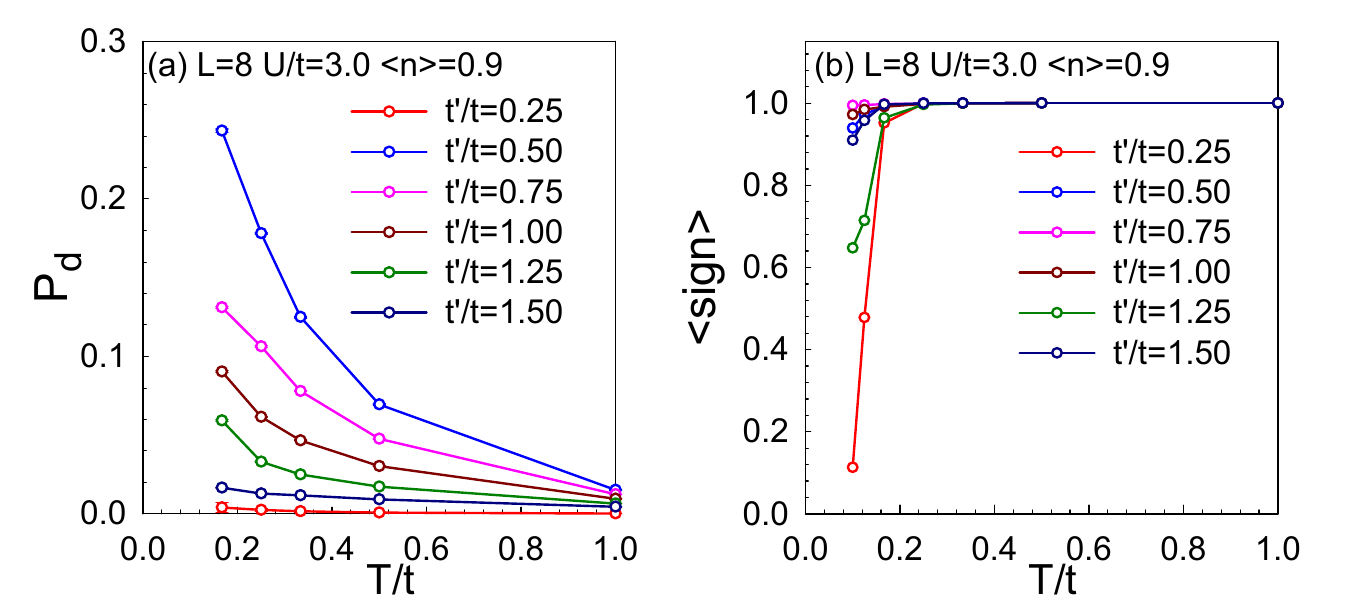}
\caption{(Color online) (a) The $d$-wave effective pairing interaction $\rm{P}_d$ as a function of temperature for different $t^{\prime}/t$
at $U/t= 3.0$, with $\avg{n}=0.9$ on a $2\times8^2$ lattice.
(b) The average sign as a function of temperature $T$ for different $t^{\prime}/t$ at $U = 3.0t$ and $\avg{n}=0.9$ on a $2\times8^2$ lattice.}
\label{Fig8}
\end{figure}

\begin{figure}[tbp]
\includegraphics[scale=0.45]{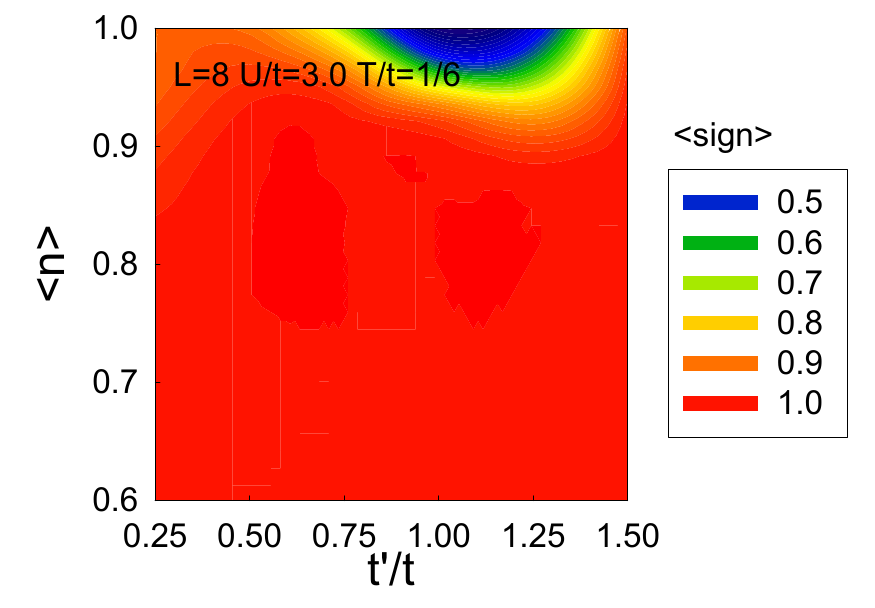}
\caption{(Color online) The average sign as a function of electron filling $\avg{n}$ for different $t^{\prime}/t$.}
\label{Fig9}
\end{figure}

We further  illustrate the temperature dependence of $d$-wave effective pairing interaction $\rm{P}_d$ for different $t^{\prime}/t$ at 
fixed electron filling
$\avg{n}=0.9$ and $U/t=3.0$ in Fig. \ref{Fig8}(a). 
Clearly, as $t^{\prime}/t$ increases, 
the value of the $d$-wave effective pairing interaction $\rm{P}_d$ decreases.
This is also evidence that the $d$-wave pairing tendency may be suppressed at $t^{\prime}/t=1.5$. 
To clarify which parameter region is accessible and reliable, we simulate the average sign decaying exponentially with
decreasing temperature for different $t^{\prime}/t$ in Fig. \ref{Fig8}(b), 
and we also plot $\langle sign \rangle$ in the $t'/t$ and 
$\avg{n}$ plane in Fig. \ref{Fig9} with $L=8$, $U/t=3.0$, $T/t=1/6$.

\section{Conclusions}

To summarize, we study the spin susceptibility, the effective pairing interaction and 
superconducting instability  of different pairing symmetries in the Hubbard model on a checkerboard lattice by using the DQMC method.
The checkerboard lattice
provides a fertile playground to study strongly
correlated physics.
We identify the dominant pairing symmetry in this system
and reveal the competition between the $d$-wave and $d$+$is$ wave
in the parameter space of electron filling $\avg{n}$ and frustration control parameter $t^{\prime}/t$.
It is found that the spin susceptibility and the effective pairing interaction 
are enhanced as the on-site Coulomb interaction increases,
demonstrating that the pairing symmetry is driven by a strong electron--electron correlation.
We also evaluate the sign problem to ensure the reliability and accuracy of our results.
Our work provides further understanding of unconventional superconductivity.

\section*{appendix}
\renewcommand{\theequation}{A\arabic{equation}}
\renewcommand{\thefigure}{A\arabic{figure}}
\renewcommand{\thesubsection}{A\arabic{subsection}}
\setcounter{equation}{0}
\setcounter{figure}{0}

\begin{figure}[h]
\includegraphics[scale=0.45]{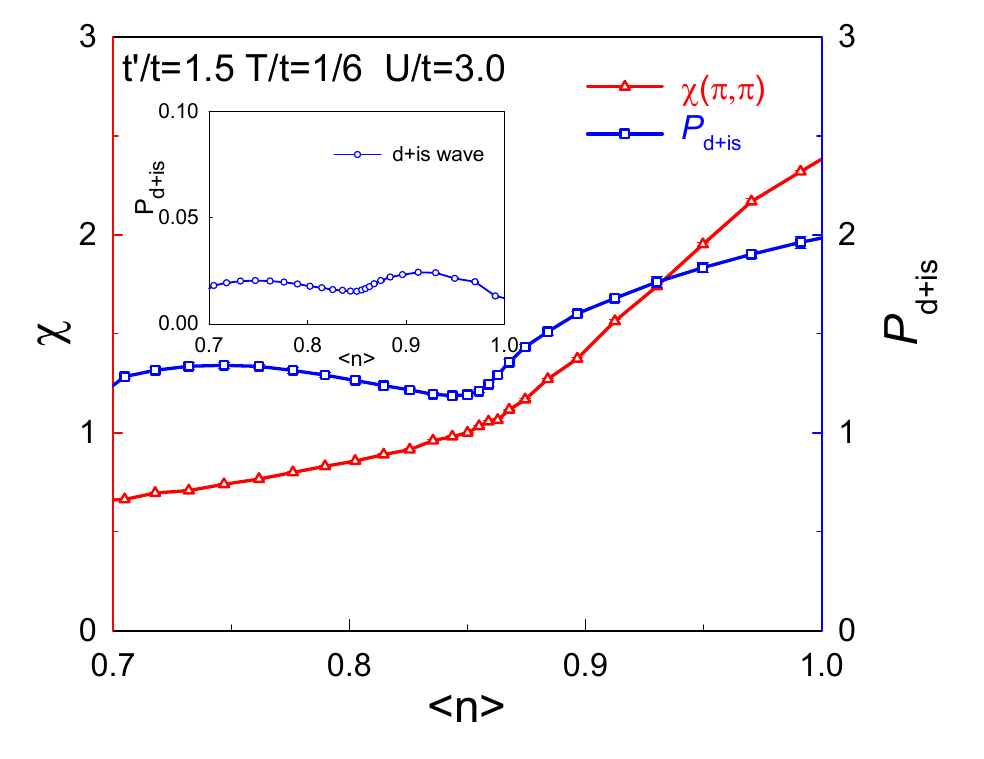}
\caption{ (Color online) The $d+is$ wave pairing susceptibility $P_{d+is}$ 
and magnetic susceptibility $\chi(\pi,\pi)$ as a function of filling at $t^{\prime}/t=1.5$, 
$T/t=1/6$, $U/t=3.0$ on a $2\times8^2$ lattice. Inset: The $d+is$ wave effective pairing interaction $\rm{P}_{d+is}$ 
as a function of filling.}
\label{FigA1}
\end{figure}

\begin{center}
\textbf{1. Intertwining of magnetism and superconductivity }
\end{center} 
In Fig.\ref{FigA1}, we studied the filling dependence of $d+is$ wave pairing susceptibility $P_{d+is}$ 
and magnetic susceptibility $\chi(\pi,\pi)$ at $t^{\prime}/t=1.5$, 
$T/t=1/6$, $U/t=3.0$ on a $2\times8^2$ lattice.
Fig.\ref{FigA1} reveals the intertwining of magnetism and superconductivity,
when magnetic susceptibility decreases, the $d+is$ wave susceptibility also decreases, 
but the effective pairing interaction increases.
Recalling the Fermi surface at $t^{\prime}/t=1.5$ in the main text,
a second band emerges and there is a potential for nesting with the other band. This will enhance magnetism near
the filling of perfect nesting at $\langle n \rangle \approx 1.0$. When the system is slightly 
deviated from perfect nesting, the superconductivity will emerge, 
which is indicated by the effective pairing interaction in the inset. 
At larger doping, $d$-wave is overwhelming and $d+is$-wave is suppressed. 

\begin{figure}[h]
\includegraphics[scale=0.35]{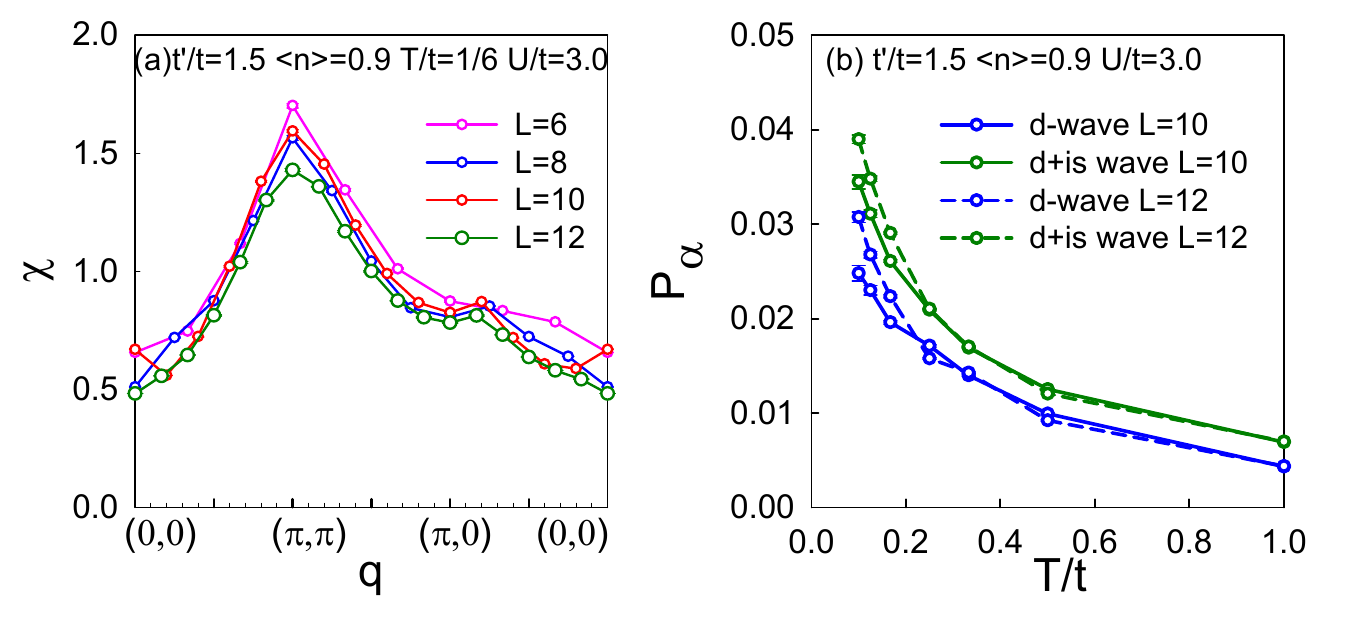}
\caption{(Color online) 
(a) The magnetic susceptibility versus momentum $q$ at different values of lattice size $L$ with 
$t^{\prime}/t = 1.5$, $\langle n \rangle = 0.9$, $T/t=1/6$ and 
$U/t = 3.0$. (b) The effective pairing interaction at different lattice size $L$ at $t^{\prime}/t = 1.5$, 
$U/t = 3.0$ and $\langle n \rangle = 0.9$.}
\label{FigA2}
\end{figure}

\begin{center}
\textbf{2. Different lattice sizes}
\end{center} 
We also check the results at different lattice sizes
at $t^{\prime}/t = 1.5$, 
$U/t = 3.0$ and $\langle n \rangle = 0.9$ in Fig.~\ref{FigA2}. 
In Fig.~\ref{FigA2}(a),
we show the magnetic susceptibility 
for different lattice sizes $L$ at $T/t = 1/6$.
One can see that the magnetic susceptibilities decreases as 
the lattice sizes $L$ increases.
In Fig.~\ref{FigA2}(b),
we plot the effective pairing interaction $\rm{P}_\alpha$ of $d$-wave and $d+is$ wave pairing symmetries as a function of temperature.
We find the results are consistent
at different lattice sizes $L$, namely, the dominant pairing symmetry is $d+is$ wave, even the $d+is$ wave effective pairing interaction
increases as $L$ increases.

\begin{figure}[ht]
\includegraphics[scale=0.36]{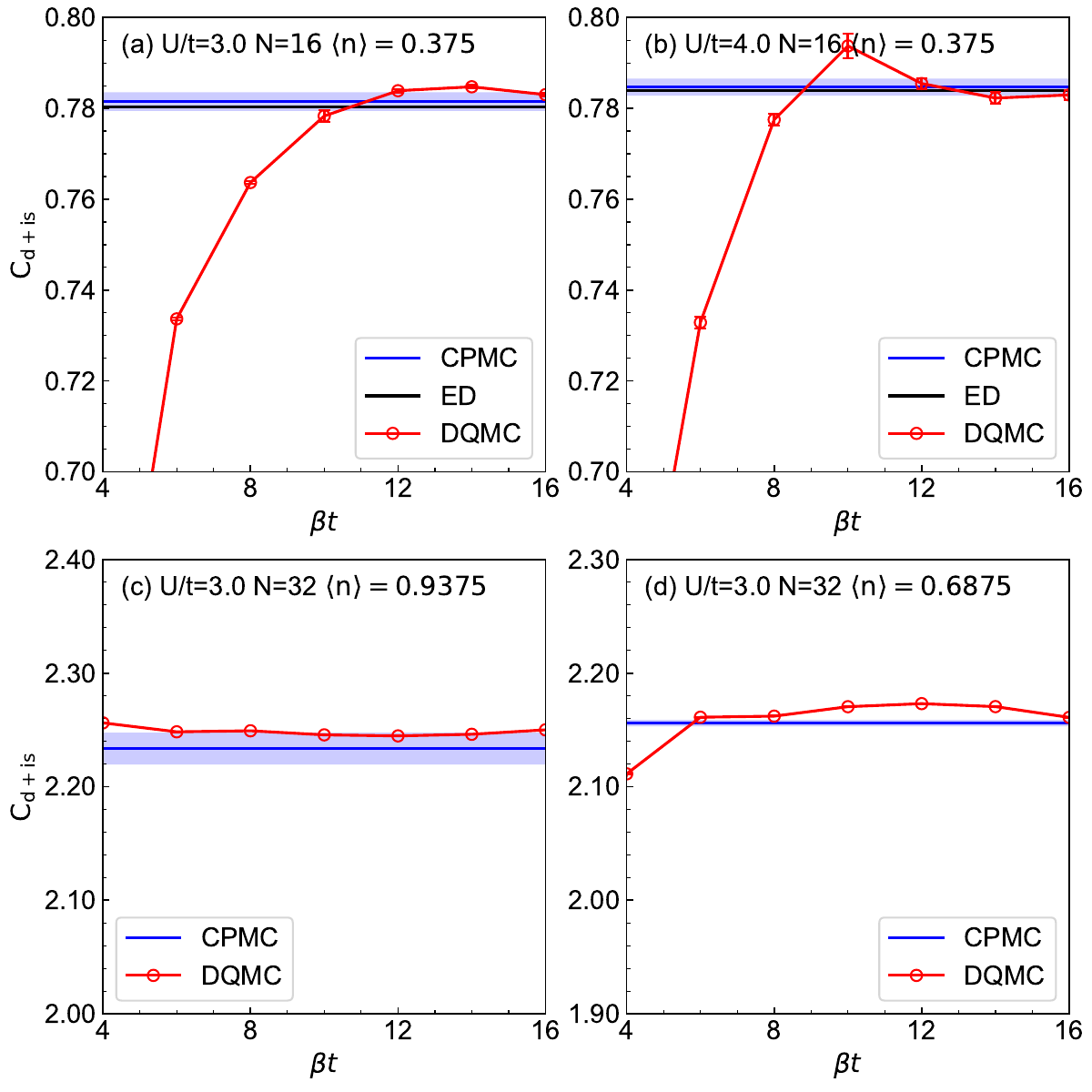}
\caption{(Color online) 
(a) On a 16-sites checkerboard lattice, the pairing correlation as a function of inverse temperature $\beta$ by DQMC method (red) is compared with
 ground sate pairing correlation by ED (dark) and CPMC method (blue), where shaded area indicates the statistical error of the CPMC results.
The parameters used are $U/t=3.0$, $t^{\prime}/t=1.5$ and $\langle n \rangle = 0.375$.
(b) The same as (a) but $U/t=4.0$.
(c) The results on a 32-sites checkerboard lattice by DQMC (red) and CPMC (blue with shaded area) method at $t^{\prime}/t=1.5$, $U/t=3.0$, $\avg{n}=0.9375$.
(d) The same as (c) but $\avg{n}=0.6875$.
  }
\label{ed}
\end{figure}

\begin{figure}[ht]
\includegraphics[scale=0.36]{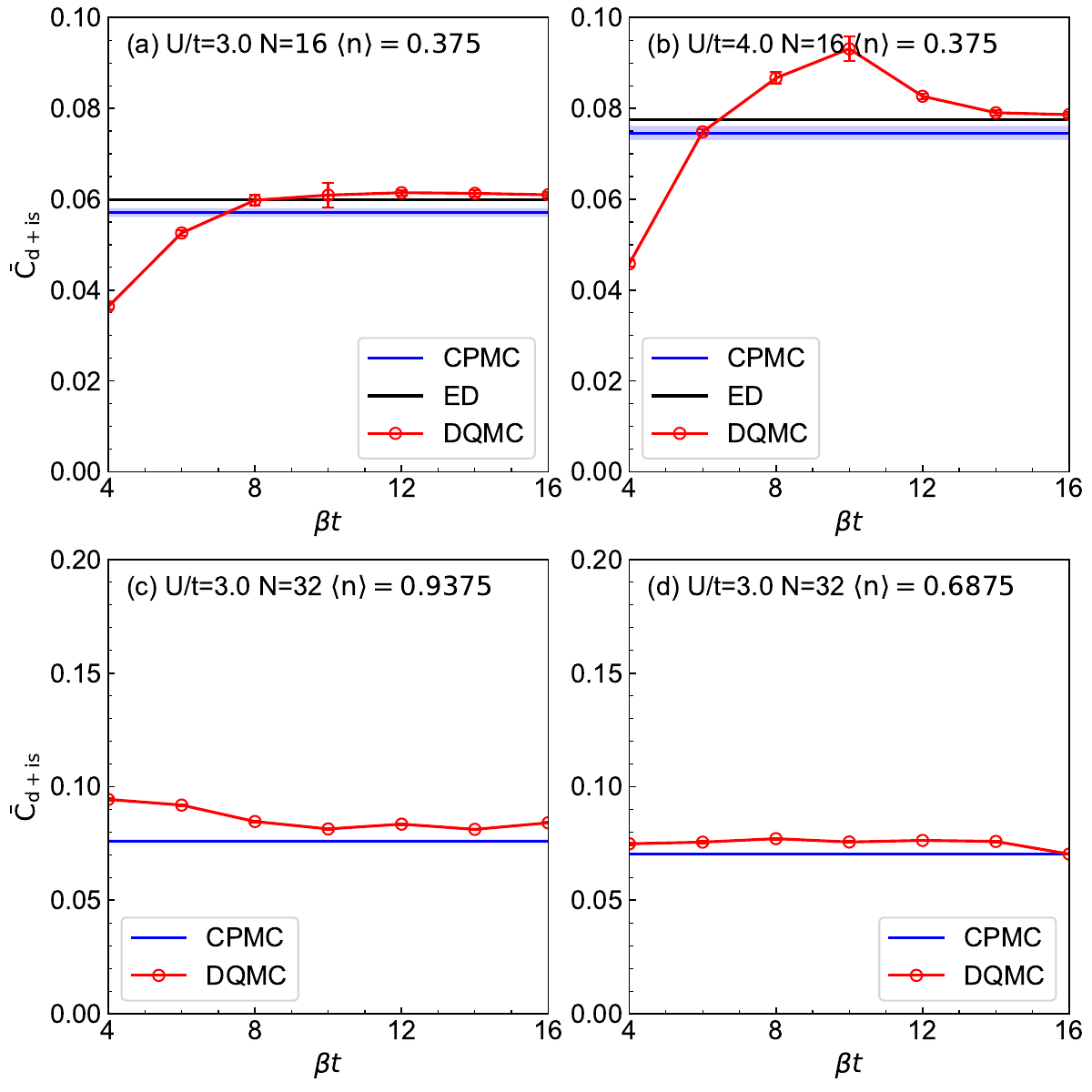}
\caption{(Color online) 
(a) On a 16-sites checkerboard lattice, the effective pairing interaction as a function of inverse temperature $\beta$ by DQMC method (red) is compared with that by ED (dark) and CPMC method (blue), where shaded area indicates the statistical error of the CPMC results.
The parameters used are $U/t=3.0$, $t^{\prime}/t=1.5$ and $\langle n \rangle = 0.375$.
(b) The same as (a) but $U/t=4.0$.
(c) The results on a 32-sites checkerboard lattice by DQMC (red) and CPMC (blue with shaded area) method at $t^{\prime}/t=1.5$, $U/t=3.0$, $\avg{n}=0.9375$.
(d) The same as (c) but $\avg{n}=0.6875$.
  }
\label{ed_V}
\end{figure}

\begin{center}
\textbf{3. Benchmark }
\end{center} 
We provide benchmark calculations of pairing correlations and effective pairing interaction, including exact diagonalization(ED) and constraint-path Quantum Monte Carlo(CPMC),
to ensure the reliability of our results by the determinant Quantum Monte Carlo method. 
As we all know, it is hard to perform ED simulations on a lattice larger than 16-sites. 
Firstly, we verified the accuracy of the CPMC algorithm and DQMC method with the ED results on a lattice with 16 sites
in Fig.~\ref{ed}(a) and (b), as well as the effective pairing interaction in Fig.~\ref{ed_V}(a) and (b).
One can see that, the ground state pairing correlation and effective pairing interaction of ED and CPMC agree with each other rather well. 
As the temperature is lowering, the results of DQMC tend to be saturated and are very near to the ground state data of ED for both $U/t=3.0$ and $U/t=4.0$.
For a larger lattice, it is impossible to perform ED simulations. 
The CPMC method has yielded very accurate results for the ground-state energy and other ground-state observables
for various strongly correlated lattice models and here we use CPMC method as a benchmark for larger lattice\cite{PhysRevB.106.134513}. 
In Fig.~\ref{ed}(c) and (d), as well as Fig.~\ref{ed_V}(c) and (d), we perform simulations on a lattice with 32 sites. 
It is found that the DQMC results are very near to the CPMC results as the inverse temperature is lowered than $\beta t=8$ at high electron filling for both 
pairing correlation and the effective pairing interaction, which indicates the reliability of our DQMC results.

\noindent
\underline{\it Acknowledgment} ---
This work was supported by Beijing Natural Science
Foundation (No. 1242022) and
Guangxi Key Laboratory of Precision Navigation Technology and Application,
Guilin University of Electronic Technology (No. DH202322).
The numerical simulations in this work were performed at the HSCC of Beijing Normal University.
\bibliography{reference}

\end{document}